%% file: master_document.tex
\documentclass[journal]{IEEEtran}
\newcommand\EatDot[1]{}
\ifCLASSINFOpdf
\else
   \usepackage[dvips]{graphicx}
\fi
\usepackage{url}

\usepackage{graphicx}
\usepackage{multirow}
\usepackage{amsmath,amssymb}
\usepackage{booktabs}
\usepackage{multirow}
\usepackage{makecell} 
\usepackage{float}

\begin{document}

\title{Enhancing Few-shot Keyword Spotting Performance through Pre-Trained Self-supervised Speech Models}

\author{Alican Gok, Oguzhan Buyuksolak, Osman Erman Okman, Murat Saraclar, \IEEEmembership{Senior Member, IEEE}
\thanks{This work has been submitted to the IEEE for possible publication. Copyright may be transferred without notice, after which this version may no longer be accessible}
\thanks{Alican Gok is both with the Department of Electrical Eng., Bogazici University and Analog Devices, Istanbul, Turkey (e-mail: alican.gok@std.bogazici.edu.tr). O. Erman Okman and Oguzhan Buyuksolak are with Analog Devices, Istanbul, Turkey. Murat Saraclar is with the Department of Electrical Eng., Bogazici University, Istanbul, Turkey}
}

\markboth{Submitted to IEEE Signal Processing Letters}
{Shell \MakeLowercase{\textit{et al.}}: Bare Demo of IEEEtran.cls for IEEE Journals}
\maketitle

\input{abstract.tex}

\IEEEpeerreviewmaketitle

\input{introduction.tex}

\input{methodology.tex}

\input{results.tex}

\input{conclusion.tex}

\input{acknowledgment.tex}

\bibliographystyle{ieeetr}
\bibliography{references}

\end{document}

%% file: abstract.tex
\begin{abstract}
Keyword Spotting plays a critical role in enabling hands-free interaction for battery-powered edge devices. Few-Shot Keyword Spotting (FS-KWS) addresses the scalability and adaptability challenges of traditional systems by enabling recognition of custom keywords with only a few examples. However, existing FS-KWS systems achieve subpar accuracy at desirable false acceptance rates, particularly in resource-constrained edge environments. To address these issues, we propose a training scheme that leverages self-supervised learning models for robust feature extraction, dimensionality reduction, and a teacher-student framework for knowledge distillation. The teacher model, based on Wav2Vec 2.0 is trained using Sub-center ArcFace loss, which enhances inter-class separability and intra-class compactness. To enable efficient deployment on edge devices, we introduce attention-based dimensionality reduction and train a standard lightweight ResNet15 student model. We evaluate the proposed approach on the English portion of the Multilingual Spoken Words Corpus (MSWC) and the Google Speech Commands (GSC) datasets. Notably, the proposed training method improves the 10-shot classification accuracy from 33.6\% to 76.9\% on 11 classes at 1\% false alarm accuracy on the GSC dataset, thus making it significantly better-suited for a real use case scenario.
\end{abstract}


%% file: introduction.tex
\section{Introduction}

\IEEEPARstart{K}{eyword} Spotting (KWS) enables hands-free interaction with devices by detecting specific spoken commands or wake-words in audio streams. Traditional KWS systems are limited in adaptability and scalability, as they require thousands of training examples for a fixed vocabulary set, and require computation and memory resources beyond the capabilities of embedded \textit{edge} devices, often battery powered smart sensors. Few-shot keyword spotting systems (FS-KWS) address these challenges by enabling the recognition of custom keywords from only a handful of examples, using deep neural networks up to only a few hundred thousand parameters.

FS-KWS leverages metric learning frameworks allowing the model to readily adapt to new classes (i.e. keywords) by enrolling only a few examples \cite{FewShotMazumderIntro}. Notably, the approach of prototypical networks is to train a representation model that processes speech segments into fixed dimensional feature vectors, also referred to as embeddings \cite{FewShotPrototypical}. The goal is to learn an embedding space in which samples of the same keyword cluster together, while those of different keyword classes are well separated. This is achieved by metric learning objectives such as Triplet loss or Prototypical Loss \cite{DummyProto_Qualcommguys}. For systems that support the open-set setting, users provide a few spoken examples of their keywords during the enrollment phase, from which class prototypes are computed as the mean of the embeddings \cite{UnsupervisedFewShotKWS}. The inference decisions are based on the shortest distance between the embedding of the test samples and the enrolled class prototypes. Unfortunately, the leading approaches in this new field achieve low accuracies at desired false acceptance rates \cite{RusciMicro, RusciInterspeech}.

A common approach to boost the performance of supervised speech tasks is to incorporate self-supervised learning (SSL) models, such as Wav2Vec 2.0 \cite{neurips2020_wav2vec_2} or HuBERT \cite{hubert}, which generate rich frame-level features. These SSL models output a feature vector for each frame of speech, resulting in many vectors per the duration of a keyword. Methods such as mean-pooling \cite{SUPERB}, convolutional encoders \cite{wav2kws}, or attention-based extraction \cite{ExtractingTokensfromSSLModels} have been used to obtain lower dimensional embeddings for KWS and other discriminative tasks. 

\begin{figure}
\centerline{\includegraphics[width=\columnwidth]{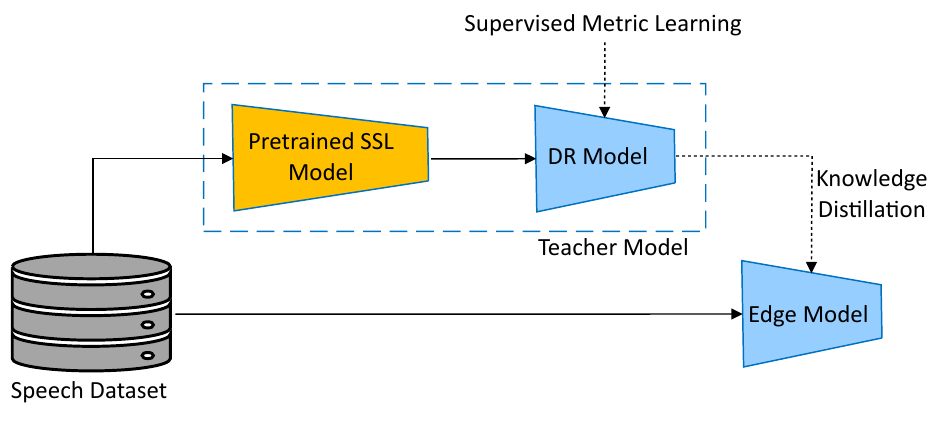}}
\caption{Proposed approach for edge FS-KWS model training}
\label{overall_system}
\end{figure}

In this paper, we propose a FS-KWS model training scheme as shown in Fig.~\ref{overall_system} that combines a pretrained SSL model with an alternative metric learning objective. Specifically, our training scheme leverages knowledge distillation, with the SSL-based teacher network optimized by Sub-center ArcFace (SCAF) loss ~\cite{SubcenterArcFace}, as opposed to triplet loss variants. ArcFace ~\cite{ArcFace}, originally developed for face recognition, enforces class separation by adding an angular margin to the softmax classifier. SCAF further enhances this by introducing multiple sub-centers per class and selecting the closest one for each sample, better capturing intra-class variation. These sub-centers allow the model to represent diverse samples within the same class more effectively. ArcFace has recently been applied to audio-related tasks such as emotion recognition~\cite{ArcFaceforEmotionRecognition} and speaker verification~\cite{ArcfaceForSpeakerVerification}. To the best of our knowledge, our work is the first to employ SCAF in the context of audio discrimination for FS-KWS. We further explore training strategies for edge-friendly student models, and benchmark our approach against the leading FS-KWS method on public datasets.

The main contributions of this work are as follows:
\begin{itemize}
    \item Introduction of a novel FS-KWS framework that enhances keyword detection capabilities on edge devices.
    \item Utilization of SCAF for efficient dimensionality reduction of embeddings from a pre-trained SSL model.
    \item Demonstrating the effectiveness of the proposed framework against the leading method in training edge-friendly FS-KWS on the MSWC and GSC datasets. 
\end{itemize}

%% file: methodology.tex
\section{Teacher-Student Representation Learning}

In this section, we introduce an approach for training a teacher model that generates edge-compatible discriminative embeddings. Following this, we outline the methodology for training a lightweight student model that effectively utilizes these embeddings.

\subsection{The Teacher Representation Model}

Self-supervised learning (SSL) models have proven to be highly effective in generating rich embeddings for various audio processing tasks. However, the inherent high dimensionality of these embeddings poses significant challenges for implementation, particularly on resource-constrained edge devices. To mitigate this issue, we propose a dimensionality reduction framework that not only compresses SSL embeddings but also enhances their discriminative capabilities for keyword spotting applications.

Our approach is built upon the Wav2Vec 2.0 architecture \cite{neurips2020_wav2vec_2}, a leading SSL model developed for speech recognition pipelines. Our feature extraction process involves deriving high-dimensional embeddings from the 16th transformer layer of a pre-trained Wav2Vec 2.0 model. This specific layer was selected based on extensive empirical analysis in \cite{WordEmbeddingsfromSSLmodels}.

\begin{figure}
\centerline{\includegraphics[width=\columnwidth]{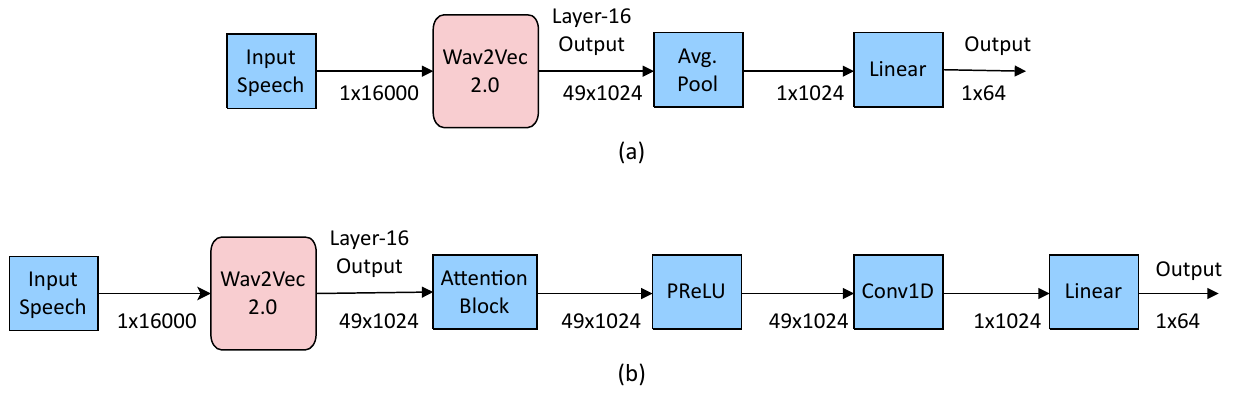}}
\caption{Proposed Teacher Representation Model with (a) simple pooling encoder, (b) attention encoder dimensionality reduction models}
\label{teacher_model}
\end{figure}

Given that Wav2Vec 2.0 operates with an approximate 20 ms stride, resulting in 49×1024 dimensional features for a 1-second utterance, processing such a volume of data on low-power edge processors is not feasible. Our proposed framework addresses this challenge by employing one of the following two distinct dimensionality reduction (DR) architectures to yield 64-dimensional embeddings.

\textit{1) Simple Pooling Encoder} given in Fig. \ref{teacher_model}(a) implements a common dimensionality reduction approach, where temporal average pooling is followed by linear projection to obtain lower-dimensional representations.

\textit{2) Attention Encoder} given in Fig. \ref{teacher_model}(b) incorporates a Scaled Dot Product Attention \cite{vaswani2017attention} mechanism followed by parametric rectified linear unit (PReLU) \cite{PReLU} activation and a one-dimensional convolutional layer (Conv1D). The attention mechanism computes temporal relationships across the input sequence, enabling the model to focus on salient temporal features. The subsequent Conv1D layer summarizes the temporal dimension to a single value through weighted averaging, effectively preserving critical temporal information while reducing computational complexity. 

The final stage of both architectures employs a linear transformation layer to further compress the feature representation, producing compact embeddings suitable for deployment on edge devices. This architectural design ensures the preservation of discriminative features while achieving significant reduction in dimensionality, which makes it particularly suitable for resource-constrained environments.

\begin{figure*}
\centerline{\includegraphics[width=\textwidth]{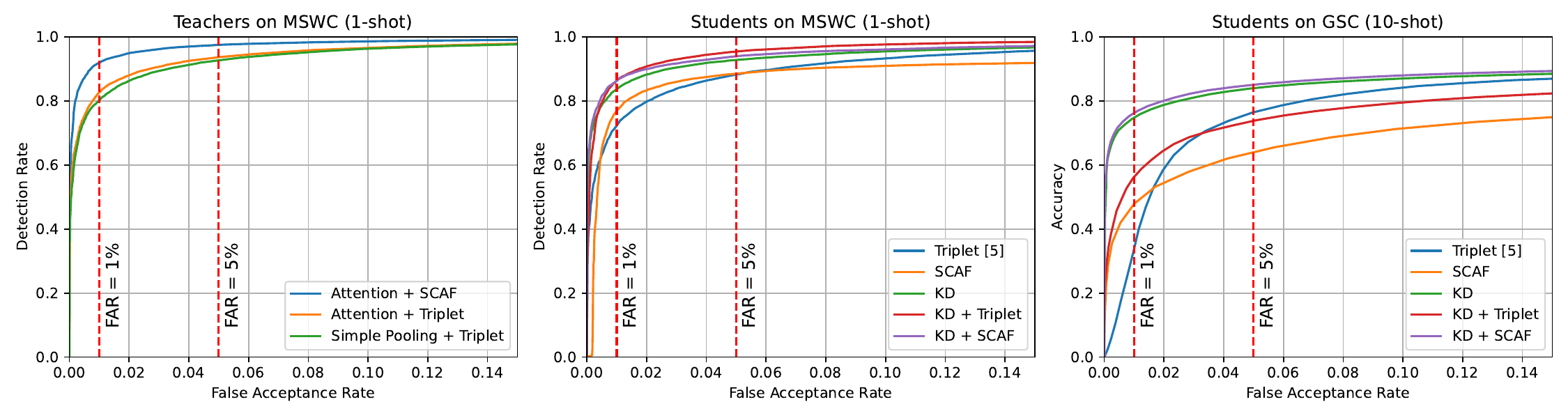}}
\caption{Performance of (a) Teacher models on the MSWC dataset under 1-shot setting for different dimensionality reduction architectures (b) Student models on the MSWC dataset under 1-shot setting (c) Student models on the GSC dataset under 10-shot setting}
\label{all_results}
\end{figure*}

These dimensionality reduction models are trained using labeled samples with two distinct loss functions: (i) Triplet ($\mathcal{L}_{\text{triplet}}$) and (ii) SCAF ($\mathcal{L}_{\text{subcenter}}$). The experiments carried out to determine the optimal dimensionality reduction model and the loss function are provided in Section \ref{sec:Experiments}. Owing to space limitations, we refer interested readers to Section II of the supplementary material for definitions of the loss functions.

\subsection{Edge-friendly Student Representation Model}

ResNet-based models are widely utilized as benchmarks in edge-focused keyword spotting (KWS) approaches \cite{SelfLearningPersonalizedKWS, DeepResidualLearningSmallFootprint, LearningEffRepKWSTriplet}. Specifically, the ResNet15 model has demonstrated superior performance over edge-friendly candidates in \cite{RusciMicro}. Our experiments therefore focus on ResNet15, which takes mel-frequency cepstral coefficients (MFCC) as input to further reduce data size, memory and computation, making it ideal for edge hardware. Compared to the teacher representation model, the student ResNet15 has three orders of magnitude fewer parameters ($480$k vs. $218$M) and computational requirements ($235$M multiply-accumulate operations vs. $63.3$G).

Knowledge distillation (KD) is an effective method for transferring knowledge between neural network models \cite{KnowledgeDistillation_Hinton}. In this technique, a pretrained teacher model guides the training of a target student model. The training process incorporates a combined loss function, which includes a distillation loss ($\mathcal{L}_{{KD}}$) and a task-specific loss ($\mathcal{L}_{{T}}$), expressed as:

\begin{equation}
\label{Eq2}
\mathcal{L} = \mathcal{L}_{{KD}} + \lambda\mathcal{L}_{{T}} 
\end{equation}

In this work, we employ mean squared error (MSE) as the distillation loss to align the embeddings of the teacher and student models. While the distillation loss by itself results in decent student models, we further explore optimizing the models via a task-specific loss in Section \ref{sec:Experiments}, for which we consider the same two alternatives as in training the teacher network: Triplet ($\mathcal{L}_{\text{triplet}}$) and SCAF ($\mathcal{L}_{\text{subcenter}}$).

%% file: results.tex
\section{Experiments}\label{sec:Experiments}

\subsection{Datasets}

To train the dimensionality reduction network of the teacher, and the edge-friendly student network, we use the train split of the English portion of Multilingual Spoken Words Corpus (MSWC) \cite{MSWCdataset}, which contains 5.5 million one-second samples, belonging to 39,000 unique words. We test our approach using the test split of the MSWC dataset, comprised of 700k samples from 8900 unique words, in addition to the Google Speech Commands (GSC) dataset \cite{GSCdataset}, a collection of 100k one-second audio clips belonging to 35 short commands.

\subsection{Evaluation Protocol}

To evaluate the few-shot keyword spotting performance, we adopt the same testing protocol as in \cite{RusciMicro}, both for teacher and student models. During the enrollment phase, $K$ samples for each keyword is fed to the frozen representation model, and the mean of the generated embeddings of each keyword class is computed to generate class prototypes. For inference, the test sample is fed to the feature extractor, and the cosine distance between the test sample embedding and the prototypes of each keyword class is computed. If the closest of these distances is less than a threshold $T$, the test sample is assigned to the class with the nearest prototype, otherwise it is assigned to the ``others" class. By changing the threshold $T$, the tradeoff between accuracy levels and false acceptance rate (FAR) of ``others" can be analyzed for different $K$-shot scenarios. 

For experiments on the MSWC dataset, we individually consider each keyword to simulate the \textit{wake-word} scenario. In each mini-experiment on this dataset, $K$ samples from a random keyword in the test set is used to generate the class prototype, and it is tested against one positive sample from the same keyword class and one negative sample from a random keyword class. This is repeated 100000 times, which we found to provide a stable evaluation across all experiments.

For the GSC dataset, we adapt the original KWS-12 test setting \cite{GSCdataset} for FS-KWS. The task is to classify 11 classes (target keywords: \{\textit{on}, \textit{off}, \textit{left}, \textit{right}, \textit{up}, \textit{down}, \textit{go}, \textit{stop}, \textit{yes}, \textit{no}\} and silence), while ideally rejecting the 25 remaining classes as ``others". The silence class includes recordings with various background noises. It should be noted that the inclusion of the silence class makes our evaluation protocol more challenging compared to the baseline work of \cite{RusciMicro}. For enrollment, $K$ samples per keyword in the training set are used to generate the class prototypes, and the entire test set is evaluated against these. To reduce variance from the few-shot sample selection, each test is repeated 100 times and results are averaged.

\subsection{Selection of Dimensionality Reduction Model}

To identify the optimal teacher model for training the lightweight representation model, we conducted experiments utilizing two dimensionality reduction architectures alongside various loss functions. In our initial experiment, we assessed the performance of the two architectures, both of which were trained using triplet loss. Each model underwent triplet training for 3000 batches, with each batch sampling 512 triplets from randomly selected word samples. The squared normalized Euclidean distance was computed for each pair of samples. We set the triplet margin to 0.5 and the model checkpoint that minimized the loss on the validation set was selected for subsequent testing. As anticipated, the attention-based dimensionality reduction architecture yielded superior results over average pooling. Consequently, we focused our investigation on the selection of loss functions specifically for this architecture. Substituting the loss function with Subcenter ArcFace loss \cite{SubcenterArcFace} and training the dimensionality reduction architecture for 10 epochs using an angular margin of $m=28.6$, a scale factor of $s=32$, and $K=3$ subcenters per class led to the best results. Repeating training with the final hyperparameters and three different random seeds, validation loss varied by less than 0.5\%.

Fig. \ref{all_results}(a) illustrates the performance of three different models on the MSWC dataset in the 1-shot setting. The results indicate that the attention-based dimensionality reduction architecture significantly outperforms others when trained with SCAF loss. Specifically, the average inter-class cosine distance between embedding pairs on the test set increases from 0.89 to 0.93 with the inclusion of the attention encoder, and further to 0.95 when the SCAF loss is applied. Correspondingly, the average intra-class distance decreases from 0.27 to 0.26 and 0.25 respectively. The distributions of these distances are provided in Section III of the supplementary material. For completeness, we trained student models using all three models as teachers. The attention-based architecture with SCAF loss also yielded the best results for the student models. For the scope of this letter, we report results exclusively for student models trained with this teacher model.

\begin{table*}[ht!]
\centering
\caption{Comparison of teacher and student (edge) network performance across MSWC and GSC datasets for \{1, 5, 10\}-shot settings.}
\label{Table1}
\begin{tabular}{l|l|c|c c c|c c c}
\toprule
\textbf{Model} & \textbf{Objective} & \textbf{\# of supports} &
\multicolumn{3}{c|}{\textbf{MSWC}} &
\multicolumn{3}{c}{\textbf{GSC}} \\
& & & $DET_{1\%}$ & $DET_{5\%}$ & AUROC & $ACC_{1\%}$ & $ACC_{5\%}$ & AUC  \\
\midrule
Teacher & SCAF & \multirow{6}{*}{\makebox[0pt][c]{1-shot}} & 91.6 & 97.4 & 99.3 & 69.0 & 77.1 & 83.4 \\
\addlinespace[4pt]
\multirow{5}{*}{Edge (Resnet15)} & Triplet~\cite{RusciMicro} & & 72.3 & 88.3 & 97.5 & 23.7 & 52.3 & 71.1 \\
& SCAF & & 76.8 & 88.5 & 92.5 & 19.9 & 37.3 & 56.3 \\
& KD & & 83.7 & 92.7 & 98.1 & 42.9 & \textbf{59.2} & 73.0 \\
& KD + Triplet & & \textbf{86.3} & \textbf{95.4} & \textbf{98.9} & 29.5 & 47.6 & 66.5 \\
& KD + SCAF & & \textbf{86.3} & 93.9 & 98.3 & \textbf{44.0} & 59.0 & \textbf{73.7} \\
\midrule
Teacher & SCAF & \multirow{6}{*}{\makebox[0pt][c]{5-shot}} & 96.8 & 99.2 & 99.8 & 81.6 & 84.8 & 89.9 \\
\addlinespace[4pt]
\multirow{5}{*}{Edge (Resnet15)} & Triplet~\cite{RusciMicro} & & 89.8 & 97.0 & 99.3 & 32.6 & 70.6 & 84.7 \\
& SCAF & & 89.4 & 94.6 & 96.4 & 40.4 & 59.0 & 74.2 \\
& KD & & 95.1 & 97.9 & 98.1 & 69.2 & \textbf{80.5} & 88.4 \\
& KD + Triplet & & \textbf{95.4} & \textbf{98.8} & \textbf{99.6} & 53.5 & 71.1 & 83.6 \\
& KD + SCAF & & 95.1 & 98.1 & 99.4 & \textbf{69.5} & 80.2 & \textbf{88.6} \\
\midrule
Teacher & SCAF & \multirow{6}{*}{\makebox[0pt][c]{10-shot}} & 96.9 & 99.3 & 99.9 & 82.2 & 85.2 & 90.9 \\
\addlinespace[4pt]
\multirow{5}{*}{Edge (Resnet15)} & Triplet~\cite{RusciMicro} & & 91.1 & 97.6 & 99.4 & 33.6 & 76.4 & 87.5 \\
& SCAF & & 91.0 & 95.9 & 97.1 & 47.9 & 65.5 & 78.1 \\
& KD & & 95.7 & 98.2 & 99.5 & 74.4 & 83.9 & 90.2 \\ 
& KD + Triplet & & 95.5 & \textbf{98.9} & \textbf{99.7} & 55.8 & 74.0 & 86.1 \\
& KD + SCAF & & \textbf{96.1} & 98.4 & 99.5 & \textbf{76.9} & \textbf{85.2} & \textbf{91.0} \\
\bottomrule
\end{tabular}
\end{table*}

\subsection{Comparative Analysis of Edge Model Training Strategies} 

In order to systematically evaluate the training strategies, we trained the edge-friendly ResNet15 model using four approaches and compared these to the baseline method~\cite{RusciMicro}, wherein the same ResNet15 was trained with Triplet loss (\textit{Triplet}). We directly used the trained checkpoint from the authors' repository  \cite{RusciRepo} to evaluate the baseline method with our testing protocol. As in the baseline method, we use the first 10 MFCCs, computed using a frame size of 40 ms, a stride of 20 ms, and applying Hamming windowing for the input to the model. This results in a feature map of size 49x10 for each 1-second test sample.

The first strategy replaces Triplet loss in the baseline with SCAF (\textit{SCAF}). The second strategy employs knowledge distillation from the selected teacher model discussed in the previous section, with the following configurations: (1) \textit{KD}: trained solely with $\mathcal{L}_{KD}$ ($\lambda=0$); (2) \textit{KD + Triplet}: trained with both $\mathcal{L}_{KD}$ and $\mathcal{L}_{\text{triplet}}$ ($\lambda=0.03$), thereby balancing MSE and Triplet losses; (3) \textit{KD + SCAF}: trained with $\mathcal{L}_{KD}$ and $\mathcal{L}_{\text{subcenter}}$ ($\lambda=0.0003$). The selected $\lambda$ values are determined empirically. Each student model was trained three times with different seeds; validation losses and AUC values in Table~\ref{Table1} varied by less than 0.6\% in all cases.

Fig.~\ref{all_results}(b) illustrates the Detection Rate versus False Alarm Rate (FAR) for the MSWC dataset for the {1-shot} setting. Here, the \textit{KD + Triplet} configuration achieves the highest detection rates, followed by \textit{KD}, while the baseline method exhibits the poorest performance, particularly at FAR = 1\%.

Fig.~\ref{all_results}(c) shows accuracy versus FAR for the GSC dataset for the {10-shot} setting. The \textit{KD} model consistently outperforms all other configurations at FAR = 1\% and 5\%. The baseline \textit{Triplet} model ranks second at FAR = 5\% but drops sharply at lower FARs. This is primarily due to the inclusion of the silence class, as we observed in keyword-specific analysis not shown in this letter. We hypothesize that knowledge distillation from the pre-trained SSL model is key to robustness under unseen conditions.

A comprehensive evaluation of the training strategies is shown in Table~\ref{Table1}, which includes the best performing teacher model and all edge-friendly model training alternatives. On the MSWC dataset, \textit{KD + Triplet} and \textit{KD + SCAF} strategies show marginally better performance than \textit{KD} alone with higher AUC and detection rates at the given FARs. Since the training uses the train split of the MSWC dataset, it is not surprising to observe that combining KD loss with a task-specific loss can improve the performance when evaluated on the test split of same dataset. However, for the tests on the GSC dataset, the results are mixed. While the \textit{KD + Triplet} strategy performs significantly worse than \textit{KD}, \textit{KD + SCAF} is consistently better than \textit{KD} for the $ACC_{1\%}$ and $AUC$ metrics. In light of these results, for this cross-domain setting where the speaker and recording characteristics are different in the test scenario, the inclusion of task-specific losses during training should be carefully considered. While the triplet loss evidently introduces overfit on the training conditions, incorporating the SCAF loss maintains the cross-domain performance. Therefore, we recommend \textit{KD + SCAF} for practical applications requiring low false alarm rates, providing robustness in all settings.

%% file: conclusion.tex
\section{Conclusion}

This work introduced a novel few-shot keyword spotting framework that leverages pretrained self-supervised transformer models for robust detection on resource-constrained edge devices. By training an intermediate teacher model with Sub-center ArcFace loss, and a lightweight student model using knowledge distillation, our approach achieves significant improvements in inter-class discrimination in the embedding space, even for unseen classes and under cross-domain conditions. Future work will focus on extending this approach to multilingual settings, exploring alternative SSL architectures, and optimizing the deployment on ultra-low-power hardware.